\documentclass[submission, copyright, creativecommons, sharealike, noncommercial]{eptcs}

\usepackage{spacing}

\usepackage{enumitem}
\usepackage[autostyle, english=american]{csquotes}

\usepackage[labelfont=bf,textfont=it]{caption}

\usepackage{microtype}

\usepackage{mathtools, amsthm}
\usepackage{thmtools, thm-restate}

\usepackage{float}

\usepackage{mathrsfs}
\usepackage{amssymb, amsfonts}
\usepackage{textcomp}

\usepackage{centernot}

\usepackage[thicklines]{cancel}

\usepackage{stackrel}

\usepackage{tikz}

\usetikzlibrary{cd, arrows}
\usetikzlibrary{decorations.pathreplacing}

\numberwithin{equation}{section}

\usepackage{hyperref}
\usepackage[capitalise, noabbrev, nameinlink]{cleveref}

\hypersetup{
	linktoc=all,
	colorlinks=true
}

\usepackage[english]{babel}

\theoremstyle{plain}
\newtheorem{theorem}{Theorem}[section]

\theoremstyle{definition}

\newtheorem{example}[theorem]{Example}

\theoremstyle{remark}
\newtheorem{remark}[theorem]{Remark}

\usepackage{subcaption}

\usepackage{todonotes}
\usepackage{fonttable}
\usepackage{stmaryrd}
\usepackage{wasysym}
\usepackage{tabularx}
\usepackage{expl3}
\usepackage{xparse}

\usepackage{mama, quiver}

\DeclareFontFamily{U}{FdSymbolA}{}
\DeclareFontShape{U}{FdSymbolA}{m}{n}{
    <-> s * [1] FdSymbolA-Book
}{}
\DeclareFontShape{U}{FdSymbolA}{m}{b}{
    <-> s * [1] FdSymbolA-Medium
}{}
\DeclareSymbolFont{fdsymbols}{U}{FdSymbolA}{m}{n}
\SetSymbolFont{fdsymbols}{bold}{U}{FdSymbolA}{m}{b}

\DeclareMathSymbol{\smallblackdiamond}{\mathbin}{fdsymbols}{130}
\DeclareMathSymbol{\whitestar}{\mathbin}{fdsymbols}{146}

\newcommand{\pow}{\mathcal{P}}

\newcommand{\nash}{\mathbf{n}}

\newcommand{\G}{\mathcal G}
\renewcommand{\H}{\mathcal H}

\renewcommand{\L}{\mathcal L}
\newcommand{\A}{\mathcal A}

\newcommand{\loss}{\ell}

\newcommand{\GD}{\mathsf{GD}}
\newcommand{\GF}{\mathsf{GF}}

\newcommand{\Comega}{{\mathchoice
	{\rotatebox[origin=c]{180}{$\displaystyle \Omega$}}
	{\rotatebox[origin=c]{180}{$\textstyle \Omega$}}
	{\rotatebox[origin=c]{180}{$\scriptstyle \Omega$}}
	{\rotatebox[origin=c]{180}{$\scriptscriptstyle \Omega$}}
}}

\newcommand{\curr}{\mathrm{curr}}

\newcommand{\play}{\mathsf{play}}
\newcommand{\coplay}{\mathsf{coplay}}

\DeclareMathOperator{\argmax}{\mathrm{argmax}}

\newcommand{\id}{\mathrm{id}}

\newcommand{\cat}[1]{\mathcal{#1}}
\newcommand{\ncat}[1]{\mathbf{#1}}
\newcommand{\twocat}[1]{\mathbb{#1}}

\newcommand{\Cat}{\twocat{C}\ncat{at}}

\newcommand{\Set}{\ncat{Set}}

\renewcommand{\Vec}{\ncat{Vec}}

\newcommand{\Euc}{\ncat{Euc}}
\newcommand{\Smooth}{\ncat{Smooth}}

\newcommand{\op}{\mathsf{op}}

\newcommand{\iso}[1][]{\overset{#1}{\cong}}

\newcommand{\Para}{\ncat{Para}}

\newcommand{\Lens}{\ncat{Lens}}
\newcommand{\DLens}{\ncat{DLens}}

\newcommand{\comp}{\fatsemi}

\newcommand{\opticto}{\rightleftarrows}

\makeatletter
\newcommand{\superimpose}[2]{%
  {\ooalign{$#1\@firstoftwo#2$\cr\hfil$#1\@secondoftwo#2$\hfil\cr}}}
\makeatother

\newcommand{\sel}{\mathsf{sel}}

\providecommand\event{ACT 2022}
\title{
	Diegetic Representation of Feedback\\in Open Games
}
\author{
	Matteo Capucci
	\institute{University of Strathclyde}
}
\def\titlerunning{
	Diegetic Representation of Feedback in Open Games
}
\def\authorrunning{
	M. Capucci
}
\date{\today}

\begin{document}
	\maketitle

	\begin{abstract}
		We improve the framework of open games with agency \cite{capucci2021translating} by showing how the players' counterfactual analysis giving rise to Nash equilibria can be described in the dynamics of the game itself (hence \emph{diegetically}), getting rid of devices such as equilibrium predicates.
		This new approach overlaps almost completely with the way gradient-based learners \cite{cruttwell_categorical_2022} are specified and trained.
		Indeed, we show feedback propagation in games can be seen as a form of backpropagation, with a crucial difference explaining the distinctive character of the  phenomenology of non-cooperative games.
		We outline a functorial construction of arena of games, show players form a subsystem over it, and prove that their `fixpoint behaviours' are Nash equilibria.
	\end{abstract}

	\fakesection{Motivation}
In narratology, \emph{diegetic} is what exists or occurs within the world of a narrative \cite{diegetic} (such as dialog, thoughts, etc.), as opposed to \emph{extra-diegetic elements} which happens outside that world (such as voiceovers, soundtrack, etc.).
Open games represent the situations of classical game theory in a compositional and purportedly `diegetic' way, i.e.~explicitly codifying the development of the game actions and payoff distribution phases in their specification.
Hedges proposed a framework in \cite{julesthesis} which evolved first by adopting the language of lenses \cite{ghani_compositional_2018}, and then that of parametric lenses \cite{capucci2021towards} to describe the bidirectional flow of information in games.
In their last iteration \cite{capucci2021translating,capucci2021towards}, \emph{open games with agency} are defined to be given by three functions (for concreteness, we assume to work in $\Set$):
\begin{eqalign}
	\play_\G : \Omega \times X \to Y, \quad
	\coplay_\G : \Omega \times X \times R \to S \times \Comega, \quad
	\varepsilon_\G : (\Omega \to \Comega) \to P\Omega.
\end{eqalign}
The set $\Omega$ represent \emph{strategies}, $X$ and $Y$ \emph{states} of the game, while $R$ and $S$ \emph{utility} and `\emph{coutility}', respectively.
The play function has an obvious role, choosing a next state $y \in Y$ (a \emph{move}) given the current state $x \in X$ and according to a strategy $\omega \in \Omega$.
Coplay is a bit more mysterious. If we think of $S$ and $R$ as the type of utilities a player can expect to receive at the end of the game while at stage $X$ and $Y$ respectively, coplay translates between these.
Finally, $\varepsilon_\G$ is a \emph{selection function} that encodes a player's preferences: given a valuation of strategies in $\Comega$ (called \emph{costrategies} or \emph{intrinsic utility}), $\varepsilon_\G$ returns the subset of strategies with satisfactory outcome.
This data defines a parametric lens \cite{capucci2021translating}:
\begin{equation}
	\G = (\Omega,\, \Comega,\, \varepsilon_\G,\, \play_\G,\, \coplay_\G) :(X,S) \opticto (Y,R).
\end{equation}
To analyse the game $\G$, that is, to extract its Nash equilibria, we then close the game by specifying an initial state $\bar x \in X$ and a payoff function $u:Y \to R$, and then apply $\varepsilon_\G$ to the composite $\bar x \comp \G \comp u$.

Since open games have been introduced, similar models have been proposed for learners \cite{cruttwell_categorical_2022} and Bayesian reasoners \cite{smithe2021compositional, braithwaite2022},
so that a general framework has been proposed in \cite{capucci2021towards} to gather all these examples of `cybernetic systems'.\footnote{Here we call `cybernetic' systems having a distinguished part controlling the rest.}
Despite having inspired this framework, open games remain quite singular when compared to their siblings.

First of all, \textbf{their payoff dynamics lacks a well-defined role}. This shows in the way coutilities, costrategies and utilities are all different in theory but very rarely in practice, and coplay is very often simply an identity or, even worse, a discard map, which makes hard motivating the existence of a backward pass at all (see e.g.~the translation process explained in \cite{capucci2021translating}).

\begin{figure}[ht]
	\centering
	\begin{subfigure}[b]{.4\textwidth}
		\centering
		\includegraphics[width=\textwidth]{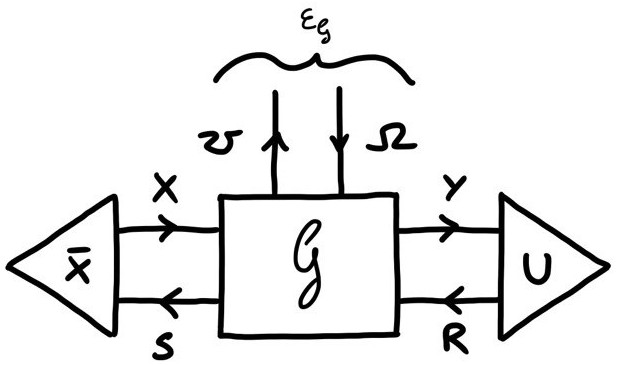}
		\fakecaption{}
		\label{fig:game}
	\end{subfigure}%
	\hspace*{10ex}
	\begin{subfigure}[b]{.4\textwidth}
		\centering
		\includegraphics[width=\textwidth]{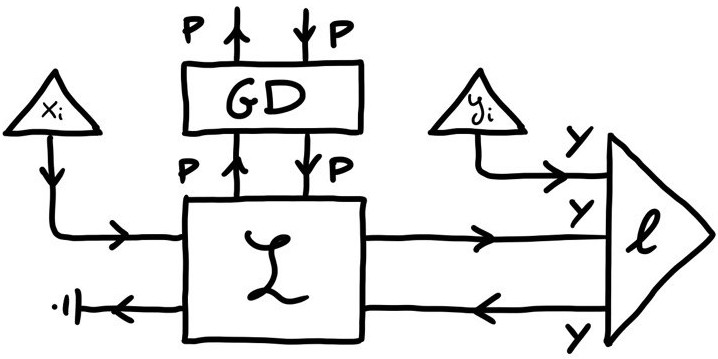}
		\fakecaption{}
		\label{fig:learner}
	\end{subfigure}%
	\vspace*{-0.65ex}
	\fakecaption{On the left, a gradient-based learner defined as in \cite{capucci2021towards,cruttwell_categorical_2022}, and on the right, an open game with agency as defined in \cite{capucci2021translating}.}
\end{figure}

Secondly, and crucially, \textbf{the dynamics they express reflect the actions happening in the game but not the game-theoretic analysis we are actually interested in}.
There's no way to know which equilibria an open game will converge to unless we pack-up the arena\footnote{The \emph{arena} of an open game with agency is the parametric lens left after forgetting about the selection function.} and then feed it to the selection function. All of this happens outside of the dynamics of the game, hence \emph{extra-diegetically}.

This issue grows into a serious conceptual flaw when we realize that according to the very notion of `system with agency' proposed by the author and his collaborators in \cite{capucci2021towards}, \emph{`open games with agency' have no agents!}
In fact, agents are supposed to be systems modelled as \emph{morphisms} plugged to the top boundary of the arena 
whereas in open games with agency players' preferences are embodied in the parameters, which are mere \emph{objects}~(\cref{fig:game}).
Contrast this with gradient-based learners (\cref{fig:learner}) where gradient descent, which implements the dynamics of an agent's learning, is explicitly represented \emph{in} the system.

\fakeparagraph{Contributions.}
In this work we correct the aforementioned problems by describing the entirety of play, payoff distribution and players' counterfactual analysis \emph{diegetically}, thus in the dynamics of the game system itself.


We achieve this by introducing two fundamental innovations.

First, we observe that \textbf{feedback propagating in an open game has to contain information about the entirety of the payoff function of the game}, hence we replace $S$ and $R$ in~\cref{fig:game} with $P^X$ and $P^Y$, where $P$ is a specified payoff object. This allows to define coplay functorially from play as precomposition with a partially-evaluated play.
This simple mechanism is enough to reproduce the information on payoffs available at each stage of a sequential or concurrent game.
Moreover, we recognize the crucial role of the lax monoidal structure of this functor, which can be blamed for the complexity of even small game-theoretic situations.

Secondly, we describe how players are embodied inside the game by their selection functions, which are now expressed as parts of a `reparameterisation' describing each player's optimization dynamic.
This fully realizes what was already intued in \cite{hedges2017higher} (`agents are their selection function') and in the drawings in \cite[§6]{capucci2021towards}, and vindicates the ideas behind open games with agency introduced in \cite{capucci2021translating}.
In fact we find out the workhorse of open games with agency, the Nash product of selection functions, decomposes in three elementary parts, the key one being `just' monoidal product of lenses.

We then show how this story shares many formal analogies with (a refinement of) gradient-based learners.
There is a formal analogy between loss covectors and payoff functions, reverse derivatives and functorially-determined coplays, `raising indices' (in the differential-geometric sense) and selection functions.
Ultimately this traces out the contours of an abstract/synthetic theory of backpropagation.


\fakeparagraph{Acknowledgements.}
We thank Jules Hedges, Philipp Zahn, Neil Ghani and Bruno Gavranovi\'c for their helpful suggestions and enthusiasm towards this work.
A special mention goes to Bruno's insistence in pointing out the conceptual flaws in the use we made of selection functions in open games with agency, as well as the numerous conversations we had together on the topic, which eventually lead me to this work.
Finally, we thank the ACT22 reviewers for their patience in reviewing an early version of this manuscript.


	\fakesection{Diegetic open games}
\label{sec:games}
We start by describing our proposed notion of \emph{diegetic open games}.
As anticipated, the key idea is to recognize that in a strategic game, players have to observe the entirety of their payoff functions with other players' actions taken into account.
This is done by fixing utility, coutility and intrinsic utility types to be of the form $P^Y$, $P^X$ and $P^\Omega$, representing entire payoff functions. Then such functions are propagated through the game in a way which is formally identical to backpropagation in learners, and thus amenable to the same mathematical treatement.
Thus $\coplay_\G$ is actually functorially determined from $\play_\G$, as a kind of reverse derivative.

\fakesubsection{Preliminaries}
Fix a finitely complete category $\cat S$.
The category $\DLens(\cat S$) of \emph{dependent lenses} over $\cat S$ has objects given by pairs of an object $Y: \cat S$ and a map $p:R \to Y$, and maps given by diagrams of the form:
\begin{diagram}[sep=4ex]
\label{diag:dep-lens}
	\textcolor{rgb,255:red,214;green,92;blue,92}{S} & \textcolor{rgb,255:red,214;green,92;blue,92}{R \times_Y X} & R \\
	X & \textcolor{rgb,255:red,214;green,92;blue,92}{X} & \textcolor{rgb,255:red,214;green,92;blue,92}{Y}
	\arrow[Rightarrow, no head, from=2-1, to=2-2]
	\arrow["f"', color={rgb,255:red,214;green,92;blue,92}, from=2-2, to=2-3]
	\arrow["p", from=1-3, to=2-3]
	\arrow["{f^*(p)}"', from=1-2, to=2-2]
	\arrow[from=1-2, to=1-3]
	\arrow[from=1-1, to=2-1]
	\arrow["{f^\sharp}"', color={rgb,255:red,214;green,92;blue,92}, from=1-2, to=1-1]
	\arrow["\lrcorner"{anchor=center, pos=0.125}, draw=none, from=1-2, to=2-3]
\end{diagram}
In the internal language of $\cat S$ \cite{shulman2021homotopy}, these maps can be denoted as $f: X \to Y$ and $f^\sharp_x : (x : X) \times R\, f(x) \to S\, x$.
The full subcategory of $\DLens(\cat S)$ spanned by those $p$ which are projections is the category of \emph{simple lenses} over $\cat S$, $\Lens(\cat S)$. The $f^\sharp$ part of simple lenses has type $X \times R \to S$.

Dependent lenses can be built from any indexed category $F:\cat S^\op \to \Cat$, in which case we denote them by $\DLens(F)$. A detailed definition and intuition is given in \cite{spivak2019generalized}.

The 2-category $\Para(\cat S)$ \cite[§2]{capucci2021towards} is the strictification of the bicategory whose objects are given by objects of $\cat S$, morphisms $X$ to $Y$ by a choice of parameter $\Omega : \cat S$ and a map $f:\Omega \times X \to Y$, and 2-morphisms $(\Omega, f) \twoto (\Xi, g) : X \to Y$ by maps $\Omega \to \Xi$ making the obvious triangle commute (see \emph{loc. cit.}, though we have reversed the direction of 2-cells here), which are called \emph{reparameterisations}.
Composition of morphisms $(\Omega,f):X \to Y$ and $(\Xi, g):Y \to Z$ is given by
\begin{equation}
	(\Xi \times \Omega, \Xi \times \Omega \times X \nlongto{\Xi \times f} \Xi \times Y \nlongto{g} Z)
\end{equation}
This makes it associative only up to coherent isomorphism, hence the strictification. Same applies to the identites, which are given by $(1, 1 \times X \nto{\pi_X} X)$.

Notice the construction of $\Para(\cat S)$ only used the cartesian monoidal structure of $\cat S$. In fact such a construction is functorial over cartesian monoidal categories. Given a lax monoidal functor \cite[Definition 1.2.14]{johnson2021} $F:\cat S \to \cat T$, with laxators $\ell_{X,Y} : F(X) \times F(Y) \to F(X \times Y)$, we get a lax 2-functor \cite[Definition 4.1.2]{johnson2021} $\Para(F) : \Para(\cat S) \to \Para(\cat T)$ defined on objects as $F$ and on a morphisms $(\Omega, f):X \to Y$ as
\begin{equation}
	\Para(F)(\Omega, f) = (F(\Omega), F(\Omega) \times F(X) \nlongto{\ell_{\Omega,X}} F(\Omega \times X) \nlongto{F(f)} Y).
\end{equation}
Since $\ell_{\Omega,X}$ is, in principle, not invertible, this means $\Para(F)$ preserves composition only up to coherent non-invertible morphism.
Explicitly, there is a reparameterisation $\Para(F)(\Omega,f) \comp \Para(F)(\Xi, g) \twoto \Para(F)((\Omega, f) \comp (\Xi, g))$, given by $\ell_{\Xi, \Omega}$. Likewise applies to preservation of identities. The well-definedness of these reparameterisations followz from the axioms of lax monoidal structure $\ell$ \cite[Diagram 1.2.14]{johnson2021}.

\fakesubsection{Building arenas}
We now describe the most simple form of games, deterministic, complete information games, with our new machinery.

Fixing a \emph{payoff object} $P$ (often $P = \R^N$, with $N$ the number of players), to a map $f:X \to Y$ we can associate the map $P^f : P^Y \to P^X$ given by precomposition with $f$.
This defines a functor $P^{(-)} : \Set \to \Set^\op$, which we can lift to a lax monoidal functor
\begin{equation}
	P^* : \Set \longto \Lens(\Set)
\end{equation}
sending $f:X \to Y$ to $(f, \pi_2 \comp P^f) : (X, P^X) \opticto (Y, P^Y)$.
Abusing notation, we'll denote by $P^*f$ both this lens and its backward part, and same with objects: $P^*X := P^X$.
Notice landing in lenses is crucial to give $P^*$ a lax monoidal structure: while its unitor $\eta : (1,1) \opticto (1,P)$, given by $(1, !_P)$ would be definable anyway; the laxator $(1_{X,Y}, \nash_{X,Y}) : (X, P^*X) \otimes (Y, P^*Y) \opticto (X \times Y, P^*(X \times Y))$, which we call \emph{Nashator}, is defined by partial evaluation at the residuals:
\begin{eqalign}
\label{eq:game-laxator}
	\nash_{X,Y} : X \times Y \times P^*(X \times Y) &\longto P^*X \times P^*Y\\
	(\bar x, \bar y, u) &\longmapsto \langle u(-, \bar y),\; u(\bar x, -) \rangle
\end{eqalign}

Ideally, this functor promotes a play function into a lens obtained by canonically adding a `coplay' function; but since play functions are actually \emph{parametric}, we need to apply $\Para$ to $P^*$ to obtain the lax 2-functor
\begin{inlinable}
	\Para(P^*) : \Para(\Set) \longto \Para(\Lens(\Set))
\end{inlinable}
so that a play function $(\Omega, \play_\G):X \to Y$ is turned into a full-blown parametric lens:
\begin{equation}
	\Para(P^*)(\Omega, \play_\G) = (\Omega, P^*\Omega,\ (1_{\Omega, X}, \nash_{\Omega, X}) \comp (\play_\G, P^*{\play_\G}))
\end{equation}
where the backward part of the right hand side boils down to
\begin{eqalign}
	\Para(P^*)(\Omega, \play_\G)^\sharp : \Omega \times X \times P^*Y &\longto P^*\Omega \times P^* X\\
	(\bar \omega, \bar x, u) &\longmapsto \langle u_\Omega,\; u_X\rangle\ \text{where}\ u_\Omega = u(\play_\G(\bar x, -))\\
	&\phantom{\ \longmapsto \langle u_\Omega,\; u_X\rangle\ \text{where}\ }
														u_X = u(\play_\G(-, \bar \omega))
\end{eqalign}
This definition is the workhorse of diegetic open games. Notice how $u_X$ encapsulates $\bar \omega$ as a fixed parameter, so that an opponent receiving such function later has that strategy fixed. Dually, $u_\Omega$ has $\bar x$ fixed so the player playing at this stage can probe $u$ by varying their own strategy but not the state the game, something determined, in turn, by other players' strategies.

\begin{remark}
	A word is due regarding the opportunity of fixing a payoff object $P$ for all games.
	This actually defeats the point of compositionality, as games with a different number of players would naturally require a different payoff object, and this without even mentioning how `dangerous' it is to allow all players to observe everybody else's payoff!
	In fact, one can develop a better version of the theory we describe here in which $\Set$ is replaced by a category of `objects with payoffs', so that we restore freedom in the payoff object we use for each game. For expositional reasons, here we stick to the simpler version in which $P$ is fixed.
\end{remark}

\begin{example}[Pure sequential game]
\label{ex:pure-seq-game}
	Consider a very simple game in which two players make one move each, in succession.
	The first player has strategy space $\Omega$ and play function $(\Omega,\play_\G ) : X \to Y$, whereas the second player has strategies $\Xi$ and play $(\Xi,\play_\H):Y \to Z$:

	\begin{figure}[h]
		\centering
		\includegraphics[width=.55\textwidth]{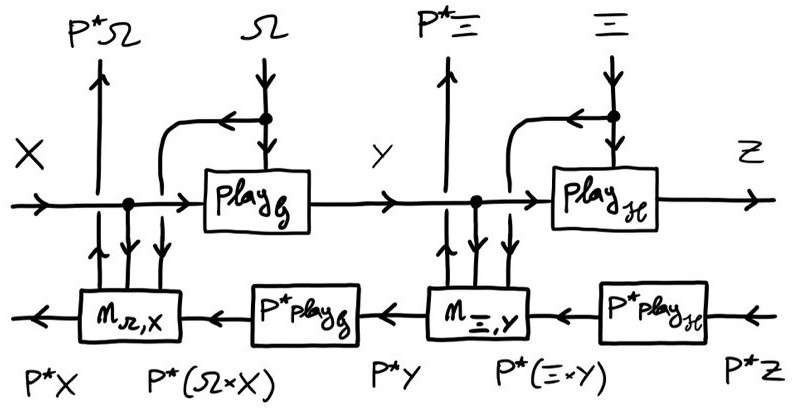}
		\fakecaption{}
		\label{fig:pure-seq-game}
	\end{figure}

	\cref{fig:pure-seq-game} depicts the parametric lens $\Para(P^*)(\Omega, \play_\G) \comp \Para(P^*)(\Xi,\play_\H)$. This is what we call the \emph{arena} of the game.

	Suppose a $\bar x \in X$ and a $u \in P^*Z$ are given, so as to close the open input horizontal wires in~\cref{fig:pure-seq-game}. These two pieces of data amount to a so-called \emph{context} for the game, and mathematically correspond to a further (trivially parameterised) lens $(\bar x, !_{P^*X}) : (1,1) \opticto (X,P^*X)$ and $(!_Z, u):(Z,P^*Z) \opticto (1,1)$.

	Then the remaining parametric lens has type $(\Xi \times \Omega, P^*\Xi \times P^*\Omega, \A) : (1,1) \opticto (1,1)$, which one can easily prove being equivalent to a function $\Xi \times \Omega \to P^*\Xi \times P^*\Omega$.
	Following $\bar x$ and $u$ around the arena, one can see what this function is given by
	\begin{eqalign}
		(\bar \xi, \bar \omega) \mapsto \langle u_\Xi,\; u_\Omega\rangle\ \text{where}\ &u_\Xi = \lambda \xi \,.\, u(\play_\H(\xi, \play_\G(\bar x, \bar \omega))\\
		&u_\Omega = \lambda \omega \,.\, u(\play_\H(\bar \xi, \play_\G(\bar x, \omega))
	\end{eqalign}
	These two functions are thus giving, to each player, all the information needed to compute their optimal strategies \emph{given the other player's strategy}. $\Para(P^*)$ makes these payoff functions emerge automatically from the information flow of lenses and from the careful use of Nashators.
\end{example}



\fakeparagraph{Payoff costates.}
As we've seen in the latter example, an arena needs, eventually, to be closed by a context.
The data of an initial state is not particularly interesting, but we need to spend a few words on the construction of \emph{payoff costates}.
Until now, open games shared the definition of payoff function with traditional strategic games: a payoff costate $(Z,P) \opticto (1,1)$ encodes exactly the information of a payoff function $Z \to P$. Now, however, a costate has to emit not just the the payoff corresponding to a given outcome of the game, but the entire payoff function.

The most direct way to do so is to have a payoff function $u:Z \to P$ being promoted to a costate $\const\,u : (Z, P^*Z) \opticto (1,1)$ in $\Lens(\Set)$ by

\begin{equation}
\label{eq:const-u}
	\const\,u = P^*u \comp (!_P, \const\, \id)
\end{equation}
where $\const\, \id : P \to P^P$ is the constant map picking the identity of $P$.
This costate effectively ignores the outcome of the game, and returns $u$ regardless.
Alternatively, if $P$ has the structure of a group, we can keep the information about the outcome and define 
\begin{equation}
	\Delta u = P^*u \comp (!_P, \curr(-))
\end{equation}
where $\curr(-) : P \to P^P$ is the curried subtraction of $P$. This effectively composes to the costate corresponding to the function
\begin{eqalign}
	\Delta u : Z &\longto P^*Z\\
			\bar z &\longmapsto \lambda z . (u(z) - u(\bar z)).
\end{eqalign}
which is a sort of `discrete differential' of $u$. Eventually this would get to players as a continuation describing their possible \emph{increment} in payoff as a function of their deviation.
In traditional game theory $\Delta u$ is known as \emph{regret} \cite[§3.2]{essentials}. We believe it to be more conceptually convicing than the constant costate, especially as we compare games with other cybernetic systems in~\cref{sec:learners}.

\fakesubsection{Adding players}
Once an arena is built, we can add players in it. At this stage, we only deal with the `vertical' part of a game, i.e.~we draw \emph{above} the arena (which constitutes the `horizontal' part of a game).
Here's where we specify how players team up, what they observe about each others' strategies and payoffs and, most importantly, how players process all this information to update their strategies.

The first thing to notice is that, since $\Para(P^*)$ is not strongly functorial, lifting the whole play function to an arena in one fell swoop versus lifting it piece by piece makes a difference in how players end up being segregated in coalitions.
In fact, if $\play_\G : X \to Y$ and $\play_\H : Y \to Z$ are parameterised by $\Omega$ and $\Xi$ respectively, then $\Para(P^*)(\play_\G \comp \play_\H)$ is parameterised by ${(\Xi \times \Omega,\ P^*(\Xi \times \Omega))}$
whereas\break
	$\Para(P^*)(\play_\G) \comp \Para(P^*)(\play_\H)$ is parameterised by $(\Xi \times \Omega,\ P^*\Omega \times P^*\Xi)$.

Effectively, \textbf{$\Para(P^*)(\play_\G \comp \play_\H)$ represents a game featuring a coalition of two players} with strategy space $\Xi \times \Omega$ (hence acting as one player), while \textbf{$\Para(P^*)(\play_\G) \comp \Para(P^*)(\play_\H)$ represents a game with two competing players}, with strategy spaces, respectively, $\Omega$ and $\Xi$.

The difference stems from the way feedback is received by players, and in their possible deviations.
In the first case, the two players can evaluate joint deviations since their feedback has type $\Xi \times \Omega \to P$.
In the second case, the two players can only evaluate unilateral deviations, because they receive two feedbacks $\Omega \to P$ and $\Xi \to P$ obtained by fixing either player's strategy.
We turn the first to the latter by reparameterising along the Nashator ${\nash_{\Xi,\Omega} : (\Xi \times \Omega, P^*\Xi \times P^*\Omega) \twoto (\Xi \times \Omega, P^*(\Xi \times \Omega))}$.
Thus, when used as a reparameterisation, \emph{the Nashator breaks down coalitions of players}.

\begin{example}[Sequential game]
	\label{ex:seq-game}
	Suppose we extend~\cref{ex:pure-seq-game} with another move by the first player (decided by the same staregy space $\Omega$, hence the copy in~\cref{fig:seq-game}.
	Contrary to the previous case, if we lifted the three play functions separately and then composed, we would have ended up splitting player one into two players: the long-range correlation between the first and third stage of the game forces us to lift the arena monolithically, as depicted in~\cref{fig:seq-game}.

	We then reparameterise along $\Delta_\Omega$ to clone the strategies of the first player into the third stage, and only then use $\nash_{\Omega, \Xi}$ to make sure players are split into two different coalitions.

	\begin{figure}[ht]
		\centering
		\includegraphics[width=.7\textwidth,angle=-90, origin=c]{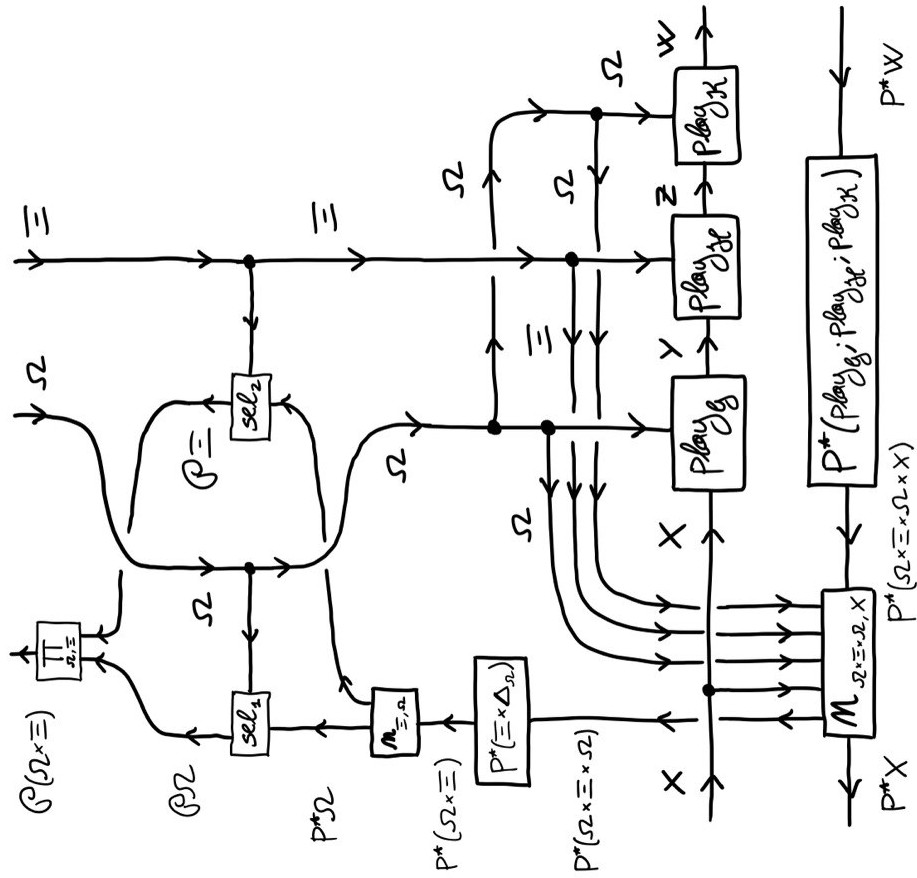}
		\fakecaption{}
		\label{fig:seq-game}
	\end{figure}
\end{example}

\begin{remark}
	Observe coalitions can always be \emph{broken} canonically, but there's no canonical way to form them.
	This is to be expected, since creating coalitions requires non\nobreakdash-canonical agreements on how to distribute payoffs among its members (so-called \emph{imputations}~\cite[Chapter~8]{essentials}).
\end{remark}

Finally, the last bit of the game specification concerns the process each player uses to turn the feedback they receive into \emph{strategic deviations}.
Usually, payoffs are numerical and players seek to maximize them. A bit more generally, players have some preferences encoded by a selection function $\varepsilon : P^*\Omega \to \pow\Omega$. We warn the reader that $P^*\Omega = P^\Omega$ is the set of $P$-valued function to $\Omega$, while $\pow\Omega$ is the powerset of $\Omega$.

A selection function fits very well in the setting we devised so far, since it has (almost) the type of the backward part of a lens $\sel : (\Omega, \pow\Omega) \opticto (\Omega, P^*{\Omega})$. We thus call such a lens a \emph{selection lens}.

\begin{remark}
	Notice the object $(\Omega, \pow\Omega)$ can be considered the `state boundary' for the player system, in the sense of \cite{myers2021book}, and betrays an implicit non-determinism in the game system. In fact, we can generalize away from sets by replacing the powerset monad $\pow:\Set \to \Set$ with other (commutative) monads, like the Giry monad on measurable spaces (yielding stochastic games) or the tangent space monad on smooth spaces (yielding differential games).
\end{remark}

\begin{remark}
	The backward part of a selection lens is actually of the form $\sel : \Omega \times P^*\Omega \to \pow\Omega$, hence a \emph{parametric selection function}.
	This suggests that $\Omega$ is even more than a set of strategies, it represents the \emph{epistemic type} of a player in the sense of Harsanyi \cite{harsanyi_games_1967}, that is, an element $\omega\in \Omega$ encodes not only the way a player plans to play but also their preferences (for instance, their aversion to risk).
	Harsanyi's games of incomplete information, at the moment codified in the framework of open games in \cite{bolt2019bayesian}, can potentially benefit a lot from the new ideas we introduced here.
\end{remark}



\fakesubsection{Games as systems}
Let's wrap up the construction we sketched so far.
The first step to specify a game is to fix the players involved ($N$) and their payoff type $P$.
The arena is built canonically from a play function $\play_\G : \Omega \times X \to Y$, where $\Omega = \Omega_1 \times \cdots \times \Omega_N$ is the product of a strategy space per player, $X$ is a type of initial states and $Y$ a type of possible final outcomes of the game.
%
Given this, we apply $\Para(P^*)$ to $\play_\G$, and get back a parametric lens $(\Omega, P^*\Omega, \A) : (X,P^*X) \opticto (Y,P^*Y)$, the arena.



\begin{remark}
	One might object that an initial state $\bar x \in X$ and a utility function $\const\,u$ (or $\Delta u$) deserve to be part of the arena too, but experience tells this data is something to provide only when we want to move on to the analysis of the game, since closing an arena prematurely hinders further composition.
	The difference between a closed and an open arena is remindful of the subtle difference between a normal (resp. extensive) \emph{form} and a normal (resp. extensive) \emph{form game}: the latter is the data of the first plus a utility function.
\end{remark}

Once the game arena has been built, we assemble the system of players over it.
Usually, such a lens will be of the form $(\bigotimes_{i=1}^N \sel_i) \comp \nash_{\Omega_1, \ldots, \Omega_N}$, where $\sel_i : (\Omega_i, \pow\Omega_i) \opticto (\Omega_i, P^*\Omega_i)$ are $N$ selection lenses.
Notice such a lens has domain $(\Omega, \pow\Omega_1 \times \cdots \times \pow\Omega_N)$,
so we precompose it with
\begin{equation}
	\prod_{i=1}^N(-) : (\Omega, \pow\Omega) \opticto (\Omega, \pow\Omega_1 \times \cdots \times \pow\Omega_N),
\end{equation}
which is the identity on the forward part and cartesian product\footnote{Or better, the canonical lax monoidal structure of the powerset endofunctor.} in the backward part (see again~\cref{fig:seq-game}).

\begin{figure}[ht]
	\centering
	\includegraphics[width=.35\textwidth]{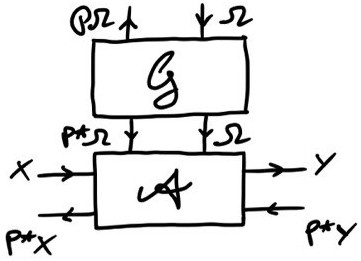}
	\fakecaption{The look of a generic diegetic open game, as a system $\G$ living over an arena $\A$.}
	\label{fig:diegetic-game}
\end{figure}

We denote the resulting lens $(\Omega, \pow \Omega) \opticto (\Omega, P^*\Omega)$ as $\G$, and this constitutes a \emph{diegetic open game} in $\Set$. Abstractly, we can consider this is a `system with boundary $\A$'~(\cref{fig:diegetic-game}), and any such system can rightfully be called a game.

%
	We stress this lens deserves to be called a system since its left (`top' in the drawings) boundary has a canonical form: the \emph{deviations} $\pow \Omega$ canonically associated to the given strategy profiles $\Omega$ (what Myers calls \emph{changes} in \cite{myers2021book}).



\fakeparagraph{Nash equilibria.}
So far, we never mentioned Nash equilibria. We have claimed that the way we have woven together the various pieces of a game reproduces, diegetically, the counterfactual analysis players do in a non-cooperative strategic game.

To see why our claim holds, let's analyze a game system constructed from a normal form $(N, \Omega)$, following the above recipe. Here $N$ is a finite set of players and $\Omega = \Omega_1 \times \cdots \times \Omega_N$.

Since normal forms dispense completely with dynamical information, the associated arena will be trivial: we set $X=1$, $Y=\Omega$\footnote{Note usually this set is called $A$ for \emph{actions}, but we prefer to keep notation consistent.} and $\play_{(N,\Omega)} := \pi_\Omega : \Omega \times 1 \to \Omega$.
Hence the arena of the game is $\A_{(N,\Omega)} = \Para({\R^N}^*)(\play_{(N,\Omega)})$.

Now we focus on players.
In a traditional non-cooperative game, they simply maximize their payoff, so that player $i$ acts according to the selection lens
\begin{equation}
	\sel_ i = (1_{\Omega_i}, \lambda (\bar\omega_i, u) . \argmax^\R u_i) : (\Omega_i, \pow\Omega_i) \opticto (\Omega_i, \R^{N \times \Omega_i}).
\end{equation}
We package this into a systems of players
\begin{equation}
	\G_{(N,\Omega)} = {\textstyle \prod_{i=1}^N}(-) \comp \left(\bigotimes_{i=1}^N \sel_i \right) \comp \nash_{\Omega_1, \ldots, \Omega_N} : (\Omega, \pow\Omega) \opticto (\Omega, \R^{N \times \Omega}).
\end{equation}
The translation of $(N,\Omega)$ is then given by the parametric lens $(\Omega, \pow \Omega, \G_{(N,\Omega)}^*\A_{(N,\Omega)}) : (1,1) \opticto (\Omega, \R^{N \times \Omega})$ obtained by plugging $\G_{(N,\Omega)}$ on the top boundary of $\A_{(N, \Omega)}$.




\begin{theorem}
\label{th:nash}
	Let $(N, \Omega=\Omega_1 \times \cdots \times \Omega_N, u:\Omega \to \R^N)$ be an $N$-players, strategic game in normal form \cite[Definition 1.2.1]{essentials}.
	Let $\G^*(\A_{(N,\Omega)} \comp \const\,u)$ be its translation to a diegetic open game, as described above, where $\const\,u$ has been in defined in~\eqref{eq:const-u}.
	Let $\G_{(\Omega,u)}: \Omega \to \pow\Omega$ be the set-valued function corresponding to such a closed parametric lens.
	Then a strategy profile $\bar\omega \in \Omega$ is a Nash equilibrium for $(\Omega, u)$ if and only if $\bar\omega \in \G_{(\Omega,u)}(\bar\omega)$.
\end{theorem}
\begin{proof}
	The set-valued function equivalent to $\G$ is obtained by following around a given strategy profile $\bar\omega \in \Omega$ along the arena, which doesn't need any other input by virtue of being closed:
	\begin{align}
		\G_{(\Omega, u)}(\bar\omega) &= \{\omega \suchthat \forall i \in N, \omega_i \in \argmax^\R(\nash_\Omega(\bar\omega, u)_i)) \} \\
		&= \{(\omega_1, \ldots, \omega_N) \suchthat \forall i \in N, \forall \omega_i' \in \Omega_i,\, u_i(\bar\omega_1, \ldots, \omega_i, \ldots \bar\omega_N) \geq u_i(\bar\omega_1, \ldots, \omega_i', \ldots \bar\omega_N) \} \notag
	\end{align}
	In other words, this is the set of best responses to the strategy profile $\bar\omega$. By definition, Nash equilibria are fixpoints of the best response function.
\end{proof}

In forthcoming work, we describe a principled, general framework to extract Nash equilibria as `behaviours' of the system $\G$ over the arena $\A$, in the style of~\cite{myers_double_2021,myers2021book}.
Specifically, we show that Nash equilibria coincide, unsurprisingly, with non-deterministic fixpoints of such systems, i.e.~simulations of the trivial game. Most importantly, from such a characterization we can automatically deduce the compositionality of equilibria which is the key strength of open games.
In other words, we can show how equilibria of a composite game can be expressed in simple terms of the equilibria of its parts.



	\fakesection{Diegetic feedback as backpropagation}
\label{sec:learners}

The conceptual story behind the diegetic representation of feedback in games is not at all specific to them.
On the contrary, it opens a window on a broader conceptual story linking the categorical description of cybernetic systems featuring a `backpropagation-like' feedback dynamics (which is most of them, notable exception being open servers \cite{videla2022lenses}).
Here we outline how gradient-based learners  \cite{cruttwell_categorical_2022} share the same abstract features, in a striking example of category theory enabling a rigorous description of a previously only informal analogy.

In gradient-based learning, a smooth function $X \to Y$ is learned by optimizing a model $f:\Omega \times X \to Y$ smoothly parameterised by the variable $\omega \in \Omega$.
Conceptually, this is only possible because differential structure leaks information about the loss $\loss : Y \times Y \to \R$ `in a neighbourhood' of $(y,f(\omega,x))$, and this can be used to evaluate which changes in parameter the learner should implement to improve.
Hence it is paramount that $\loss$ is known `locally', and not just pointwise. In practice, the value of $\loss$ at $(y, f(\omega,x))$ is not even used! Only the covector $\de_{f(\omega,x)} \loss(y,-)$ is needed.

This covector is then backpropagated across the various components of the learner until a covector on $\Omega$ is obtained.
As for games, this backpropagation mechanism is effortlessly assembled by deploying the functor
\begin{inlinable}
	T^* : \Smooth \longto \DLens(\Vec_\R)
\end{inlinable}
sending each manifold $X$ to its \emph{cotangent vector bundle} $T^*X \to X$ (the fiberwise dual of its tangent bundle) and each map $f:X \to Y$ to its reverse derivative, i.e.~pullback of covectors along $f$ \cite{spivak1973comprehensive}, naturally expressed as a dependent lens $(f, T^*f):(X,T^*X) \opticto (Y,T^*Y)$.%
\footnote{Specifically, the codomain of $T^*$ is the category of dependent lenses \cite{spivak2019generalized} obtained from the indexed category of smooth $\R$-vector bundles $\Vec_\R : \Smooth^\op \to \Cat$.}

\begin{remark}
	In \cite{cruttwell_categorical_2022}, a functor very similar to $T^*$ is obtained from the structure of \emph{reverse differential category} (RDC) on the base category, but $\Smooth$ is not such a category.
	Therefore, in \emph{ibid.} the authors confine themselves to its wide subcategory $\Euc$ of Euclidean spaces.
	In light of our findings for games, it seems that considering functors $\cat S \to \DLens(\cat S)$ splitting the view fibration to be more fundamental than reverse differential structure in the sense of \cite{cockett2020reverse}.
	Already in \cite[§4]{cockett2020reverse} and \cite[Proposition 2.12]{cruttwell_categorical_2022}, it is shown how reverse differential structures can be encoded as sections of the view fibration of lenses, with extra conditions account for the `additivity' necessary in the framework of RDCs.
	It seems reasonable, therefore, to reformulate RDCs as particularly nice instances of \emph{section of feedbacks}, dualizing that of \emph{section of changes} defined by Myers in \cite{myers_double_2021,myers2021book}.
\end{remark}

\begin{remark}
	The functor $T^*$ is strong monoidal and thus is associated to a pseudofunctor $\Para(T^*)$ that promotes a smooth parametric function straight into a backpropagating model.
	Compare this with the functor $\Para(P^*)$, whose laxity is, ultimately, the source of the many interesting phenomena in non-cooperative strategic games.
	The fact $T^*$ is \emph{not} lax is attributable to the additive structure involved in each fiber of a cotangent bundle, whereby $T^*(X \times Y) \iso T^*(X+Y)$.

	In \cite{schaefer_competitive_2019} the authors consider what amounts to a different lax monoidal structure on $T^*$, one with respect to the fiberwise tensor product of vector bundles.\footnote{This also entails replacing $\Vec_\R$ with its subfunctor of vector bundles and fiberwise linear maps.} That structure is strictly lax, like that of $P^*$. Indeed, the resulting learners behave as if they are `competing', and this is found to be better adapted for training GANs, as their game-theoretic interpretation would suggest.
\end{remark}

Once an arena $\L := \Para(T^*)(\Omega, f)$ has been defined, the dynamic of an agent (which is what really deserves the name of `learner') actually doing the learning is given by a \emph{gradient flow} lens\break$\GF : (\Omega, T\Omega) \opticto (\Omega, T^*\Omega)$ which defines a system over $\L$, by reparameterisation (as in~\cref{fig:learner}).
The backward part of such a lens is a fiberwise linear morphism $(-)^\sharp : T^*\Omega \to T\Omega$.
The most common way such a morphism arises is when $\Omega$ is endowed with a Riemannian metric $g$, in which case $(-)^\sharp$ (known as `raising indices' \cite{spivak1973comprehensive}) selects the direction of steepest ascent associated to a covector, so that $u^\sharp$ is $\argmax_{v \in T_\omega\Omega} u(v)/\|v\|_g$ for a given $u \in T^*_\omega\Omega$.

As highlighted in~\cref{table:correspondence}, $(-)^\sharp$ is formally analogous to a selection function $\sel : \Omega \times P^*\Omega \to \pow\Omega$, which indeed has the same role for games.
This is corroborated by the type signatures of $\GF$ and $\sel$, both going from an object of `states and feedbacks' to an object of `states and changes'.

\begin{table}[ht]
	\centering
	\begin{tabularx}{\textwidth}{>{\centering\arraybackslash}X|>{\centering\arraybackslash}X}
		\textbf{games} & \textbf{gradient-based learners}
		\\[0.5ex]\hline
		\begin{tabular}{@{}c@{}}
			strategies\\[-0.75ex]$\Omega$
		\end{tabular} &
		\begin{tabular}{@{}c@{}}
			parameters\\[-0.75ex]$\Omega$
		\end{tabular}
		\\[2.5ex]\hline
		\begin{tabular}{@{}c@{}}
			deviations\\[-0.75ex]$\pow\Omega$
		\end{tabular}
		&
		\begin{tabular}{@{}c@{}}
			vectors\\[-0.75ex]$T\Omega$
		\end{tabular}
		\\[2.5ex]\hline
		\begin{tabular}{@{}c@{}}
			payoff functions\\[-0.75ex]$P^*\Omega := P^\Omega$
		\end{tabular}
		&
		\begin{tabular}{@{}c@{}}
			covectors\\[-0.75ex]$T^*\Omega$
		\end{tabular}
		\\[2.5ex]\hline
		\begin{tabular}{@{}c@{}}
			precomposition\\[-0.75ex]$P^*f : X \times P^*Y \to P^*X$
		\end{tabular}
		&
		\begin{tabular}{@{}c@{}}
			reverse derivative\\[-0.75ex]$T^*\!f : f^*(T^*Y) \to T^*X$
		\end{tabular}
		\\[2.5ex]\hline
		\begin{tabular}{@{}c@{}}
			selection function\\[-0.75ex]$\sel : \Omega \times P^*\Omega \to \pow\Omega$
		\end{tabular}
		&
		\begin{tabular}{@{}c@{}}
			sharp (iso)morphism\\[-0.75ex]$(-)^\sharp : T^*\Omega \to T\Omega$ (of vector bundles over $\Omega$)
		\end{tabular}
	\end{tabularx}
	\fakecaption{}
	\label{table:correspondence}
\end{table}

What might look odd is the asymmetry between $\pow\Omega$ and $\Omega$ in the signature of $\sel$, something not present in $(-)^\sharp$. Indeed, if $\Omega$ is the set of `states' of a player, then there is a dissimilarity between $T^*\Omega$ being the set of $\R$-valuations of $T\Omega$ and $P^*\Omega$ being the set of valuations on $\pow\Omega$.
This discrepancy requires a bit more scaffolding to be explained, but intuitively it amounts to observing $T^*\Omega$ is the set of \emph{linear} valuations on $T\Omega$, an likewise, when we consider only maps $f:\pow\Omega \to P$ that satisfy $f(A) = \sum_{a \in A} f(\{a\})$, these are determined by maps $\Omega \to P$.



Let us remark on another aspect, regarding discretization of such systems. Usually learners are trained with gradient \emph{descent}, not gradient \emph{flow}, due to the evident impossibility of actually performing an infinitesimal step in the gradient direction. Thus an important role is played by the exponential map of the Riemannian manifold of parameters, since it allows to move for a definite length along a given direction.
To us, this amounts to another lens $\exp_\alpha : (\Omega, \Omega \times \Omega) \opticto (\Omega, T\Omega)$ on top of a learner, whose backward part $(\omega : \Omega) \times T_\omega\Omega \to \Omega$ is indeed given by moving for an interval of time $\alpha$ along the geodesic.
Doing this turns the differential system $\GF$ into the deterministic and discrete $\GD$ described in~\cite{cruttwell_categorical_2022}.
In fact, this can be seen as a general move from differential to discrete given by a forward Euler integration scheme, similar to what is described in~\cite{libkind_operadic_2022}.

Similarly can be done for games: the analogous structure would be that of a $\pow$-algebra.\footnote{Algebra of the $\pow$ \emph{endofunctor}, not necessarily the monad.} Concretely, this map collapses the multiple possibilities of deviations to a choice of a next strategy to `try'. This can be used to define a lens analogous to $\exp_\alpha$ that transforms a non-deterministic system into a deterministic one.

	\fakesection{Conclusions}
\label{sec:conclusions}

In this work we described a new approach to the specification of compositional games in the style of open games \cite{ghani2018compositional,capucci2021translating}. It corrects some of the conceptual shortcomings of open games with agency, and uncovers deeper analogies with gradient-based learners and, speculatively, a wider range of cybernetic systems.

The new approach provides a way to specify a game using machinery analogous to reverse-mode automatic differentiation, abstractly given by a functor $P^*:\Set \to \DLens(\Set)$.
We observed how the lax monoidal structure of such functor plays a profound role in determining the dynamics of non-cooperative games, by hiding `cooperative' information.

We have shown how classical strategic games can be naturally represented as non-deterministic systems over their arenas, systems given by the dynamics of players observing their payoffs and pondering if and how to deviate from their current strategy.
The resulting parametric lens is hence a full realization of the ideas in \cite{hedges2017higher,capucci2021translating,capucci2021towards}, and brings the framework of categorical cybernetics (born with \cite{capucci2021towards}) closer to that of categorical systems theory (detailed in \cite{myers_double_2021,myers2021book}).

\fakeparagraph{Future directions.}
The new ideas brought about in this paper are not fully formed yet. In preparing this work, three more follow-up works naturally spawned.

The first, which has already been anticipated at the end of~\cref{sec:games}, concerns laying down a proper general theory of specification and simulation of cybernetic systems, in the wake of Myers' work on dynamical systems \cite{myers_double_2021,myers2021book}. In the first place, this would allow to extract Nash equilibria from diegetic open games in a principled and compositional way, with practical implications in the way these are computed.
Secondly, using analogous tools we would then be able to talk about simulations of games and more generally of non-equilibrium trajectories of game dynamics.
Lastly, we will have in place a unifying notion of `morphism of open games', which from preliminary discussions with Hedges, seems to reproduce the most important features of those in \cite{hedges2018morphisms} and \cite{ghani2018compositional}.

The second work concerns the pure game-theoretic aspects of this new definition. Can we improve the toolset of compositional game theory by leveraging a more accurate reproduction of the dynamics involved? We believe the answer to be yes, with exciting connections to the topic of Bayesian games \cite{harsanyi_games_1967} and learning theory for games \cite{fudenberg1998theory}.

The third work is an exploration of the ideas roughly outlined in~\cref{sec:learners}, with the aim of crystallizing the analogy between learners and games.
Such an abstract theory of backpropagation would formalize the intuitive picture whereby such systems come with a notion of `type of states' on which a `type of changes', a `type of scalars' depend, which together give rise to a `type of feedbacks' obtained as valuations of the first in the latter.

	\bibliography{bibliography}

\begin{thebibliography}{10}
\providecommand{\bibitemdeclare}[2]{}
\providecommand{\surnamestart}{}
\providecommand{\surnameend}{}
\providecommand{\urlprefix}{Available at }
\providecommand{\url}[1]{\texttt{#1}}
\providecommand{\href}[2]{\texttt{#2}}
\providecommand{\urlalt}[2]{\href{#1}{#2}}
\providecommand{\doi}[1]{doi:\urlalt{https://doi.org/#1}{#1}}
\providecommand{\eprint}[1]{arXiv:\urlalt{https://arxiv.org/abs/#1}{#1}}
\providecommand{\bibinfo}[2]{#2}

\bibitemdeclare{unpublished}{bolt2019bayesian}
\bibitem{bolt2019bayesian}
\bibinfo{author}{Joe \surnamestart Bolt\surnameend}, \bibinfo{author}{Jules
  \surnamestart Hedges\surnameend} \& \bibinfo{author}{Philipp \surnamestart
  Zahn\surnameend} (\bibinfo{year}{2019}): \emph{\bibinfo{title}{Bayesian open
  games}}.
\newblock \urlprefix\url{https://arxiv.org/abs/1910.03656}.

\bibitemdeclare{unpublished}{braithwaite2022}
\bibitem{braithwaite2022}
\bibinfo{author}{Dylan \surnamestart Braithwaite\surnameend} \&
  \bibinfo{author}{Jules \surnamestart Hedges\surnameend}
  (\bibinfo{year}{2022}): \emph{\bibinfo{title}{Dependent Bayesian Lenses:
  Categories of Bidirectional Markov Kernels with Canonical Bayesian
  Inversion}}.
\newblock \urlprefix\url{https://arxiv.org/abs/2209.14728}.

\bibitemdeclare{article}{capucci2021towards}
\bibitem{capucci2021towards}
\bibinfo{author}{Matteo \surnamestart Capucci\surnameend},
  \bibinfo{author}{Bruno \surnamestart Gavranović\surnameend},
  \bibinfo{author}{Jules \surnamestart Hedges\surnameend} \&
  \bibinfo{author}{Eigil~Fjeldgren \surnamestart Rischel\surnameend}
  (\bibinfo{year}{2022}): \emph{\bibinfo{title}{Towards {Foundations} of
  {Categorical} {Cybernetics}}}.
\newblock {\slshape \bibinfo{journal}{Electronic Proceedings in Theoretical
  Computer Science}} \bibinfo{volume}{372}, pp. \bibinfo{pages}{235--248},
  \doi{10.4204/EPTCS.372.17}.
\newblock \urlprefix\url{http://arxiv.org/abs/2105.06332}.
\newblock \bibinfo{note}{ArXiv:2105.06332 [math]}.

\bibitemdeclare{article}{capucci2021translating}
\bibitem{capucci2021translating}
\bibinfo{author}{Matteo \surnamestart Capucci\surnameend},
  \bibinfo{author}{Neil \surnamestart Ghani\surnameend},
  \bibinfo{author}{Jérémy \surnamestart Ledent\surnameend} \&
  \bibinfo{author}{Fredrik~Nordvall \surnamestart Forsberg\surnameend}
  (\bibinfo{year}{2022}): \emph{\bibinfo{title}{Translating {Extensive} {Form}
  {Games} to {Open} {Games} with {Agency}}}.
\newblock {\slshape \bibinfo{journal}{Electronic Proceedings in Theoretical
  Computer Science}} \bibinfo{volume}{372}, pp. \bibinfo{pages}{221--234},
  \doi{10.4204/EPTCS.372.16}.
\newblock \urlprefix\url{http://arxiv.org/abs/2105.06763}.
\newblock \bibinfo{note}{ArXiv:2105.06763 [cs, math]}.

\bibitemdeclare{inproceedings}{cockett2020reverse}
\bibitem{cockett2020reverse}
\bibinfo{author}{Robin \surnamestart Cockett\surnameend},
  \bibinfo{author}{Geoffrey \surnamestart Cruttwell\surnameend},
  \bibinfo{author}{Jonathan \surnamestart Gallagher\surnameend},
  \bibinfo{author}{Jean-Simon~Pacaud \surnamestart Lemay\surnameend},
  \bibinfo{author}{Benjamin \surnamestart MacAdam\surnameend},
  \bibinfo{author}{Gordon \surnamestart Plotkin\surnameend} \&
  \bibinfo{author}{Dorette \surnamestart Pronk\surnameend}
  (\bibinfo{year}{2020}): \emph{\bibinfo{title}{Reverse {Derivative}
  {Categories}}}.
\newblock In: {\slshape \bibinfo{booktitle}{28th {EACSL} {Annual} {Conference}
  on {Computer} {Science} {Logic}}}, \doi{10.4230/LIPIcs.CSL.2020.18}.

\bibitemdeclare{inproceedings}{cruttwell_categorical_2022}
\bibitem{cruttwell_categorical_2022}
\bibinfo{author}{Geoffrey S.~H. \surnamestart Cruttwell\surnameend},
  \bibinfo{author}{Bruno \surnamestart Gavranović\surnameend},
  \bibinfo{author}{Neil \surnamestart Ghani\surnameend}, \bibinfo{author}{Paul
  \surnamestart Wilson\surnameend} \& \bibinfo{author}{Fabio \surnamestart
  Zanasi\surnameend} (\bibinfo{year}{2022}): \emph{\bibinfo{title}{Categorical
  {Foundations} of {Gradient}-{Based} {Learning}}}.
\newblock In \bibinfo{editor}{Ilya \surnamestart Sergey\surnameend}, editor:
  {\slshape \bibinfo{booktitle}{Programming {Languages} and {Systems}}},
  \bibinfo{series}{Lecture {Notes} in {Computer} {Science}},
  \bibinfo{publisher}{Springer International Publishing},
  \bibinfo{address}{Cham}, pp. \bibinfo{pages}{1--28},
  \doi{10.1007/978-3-030-99336-8\_1}.

\bibitemdeclare{}{diegetic}
\bibitem{diegetic}
\bibinfo{author}{Merriam-Webster.com \surnamestart Dictionary\surnameend}:
  \emph{\bibinfo{title}{Diegetic}}.
\newblock \urlprefix\url{https://www.merriam-webster.com/dictionary/diegetic}.

\bibitemdeclare{book}{fudenberg1998theory}
\bibitem{fudenberg1998theory}
\bibinfo{author}{Drew \surnamestart Fudenberg\surnameend},
  \bibinfo{author}{Fudenberg \surnamestart Drew\surnameend},
  \bibinfo{author}{David~K. \surnamestart Levine\surnameend} \&
  \bibinfo{author}{David~K. \surnamestart Levine\surnameend}
  (\bibinfo{year}{1998}): \emph{\bibinfo{title}{The theory of learning in
  games}}, \bibinfo{edition}{1} edition.
\newblock \bibinfo{volume}{2}, \bibinfo{publisher}{MIT press}.

\bibitemdeclare{inproceedings}{ghani_compositional_2018}
\bibitem{ghani_compositional_2018}
\bibinfo{author}{Neil \surnamestart Ghani\surnameend}, \bibinfo{author}{Jules
  \surnamestart Hedges\surnameend}, \bibinfo{author}{Viktor \surnamestart
  Winschel\surnameend} \& \bibinfo{author}{Philipp \surnamestart
  Zahn\surnameend} (\bibinfo{year}{2018}): \emph{\bibinfo{title}{Compositional
  game theory}}.
\newblock In: {\slshape \bibinfo{booktitle}{Proceedings of the 33rd annual
  {ACM}/{IEEE} symposium on logic in computer science}}, pp.
  \bibinfo{pages}{472--481}, \doi{10.1145/3209108.3209165}.

\bibitemdeclare{article}{ghani2018compositional}
\bibitem{ghani2018compositional}
\bibinfo{author}{Neil \surnamestart Ghani\surnameend}, \bibinfo{author}{Clemens
  \surnamestart Kupke\surnameend}, \bibinfo{author}{Alasdair \surnamestart
  Lambert\surnameend} \& \bibinfo{author}{Fredrik~Nordvall \surnamestart
  Forsberg\surnameend} (\bibinfo{year}{2018}): \emph{\bibinfo{title}{A
  compositional treatment of iterated open games}}.
\newblock {\slshape \bibinfo{journal}{Theoretical computer science}}
  \bibinfo{volume}{741}, pp. \bibinfo{pages}{48--57},
  \doi{10.1016/j.tcs.2018.05.026}.

\bibitemdeclare{article}{harsanyi_games_1967}
\bibitem{harsanyi_games_1967}
\bibinfo{author}{John~C. \surnamestart Harsanyi\surnameend}
  (\bibinfo{year}{1967}): \emph{\bibinfo{title}{Games with incomplete
  information played by “{Bayesian}” players, {I}–{III} {Part} {I}. {The}
  basic model}}.
\newblock {\slshape \bibinfo{journal}{Management science}}
  \bibinfo{volume}{14}(\bibinfo{number}{3}), pp. \bibinfo{pages}{159--182},
  \doi{10.1287/mnsc.14.3.159}.
\newblock \bibinfo{note}{Publisher: INFORMS}.

\bibitemdeclare{phdthesis}{julesthesis}
\bibitem{julesthesis}
\bibinfo{author}{Jules \surnamestart Hedges\surnameend} (\bibinfo{year}{2016}):
  \emph{\bibinfo{title}{Towards compositional game theory}}.
\newblock Ph.D. thesis, \bibinfo{school}{Queen Mary University of London}.

\bibitemdeclare{article}{hedges2018morphisms}
\bibitem{hedges2018morphisms}
\bibinfo{author}{Jules \surnamestart Hedges\surnameend} (\bibinfo{year}{2018}):
  \emph{\bibinfo{title}{Morphisms of open games}}.
\newblock {\slshape \bibinfo{journal}{Electronic Notes in Theoretical Computer
  Science}} \bibinfo{volume}{341}, pp. \bibinfo{pages}{151--177}.

\bibitemdeclare{inproceedings}{hedges2017higher}
\bibitem{hedges2017higher}
\bibinfo{author}{Jules \surnamestart Hedges\surnameend}, \bibinfo{author}{Paulo
  \surnamestart Oliva\surnameend}, \bibinfo{author}{Evguenia \surnamestart
  Shprits\surnameend}, \bibinfo{author}{Viktor \surnamestart
  Winschel\surnameend} \& \bibinfo{author}{Philipp \surnamestart
  Zahn\surnameend} (\bibinfo{year}{2017}): \emph{\bibinfo{title}{Higher-order
  decision theory}}.
\newblock In: {\slshape \bibinfo{booktitle}{International Conference on
  Algorithmic Decision Theory}}, \bibinfo{organization}{Springer}, pp.
  \bibinfo{pages}{241--254}, \doi{10.1007/978-3-319-67504-6\_17}.

\bibitemdeclare{book}{johnson2021}
\bibitem{johnson2021}
\bibinfo{author}{Niles \surnamestart Johnson\surnameend} \&
  \bibinfo{author}{Donald \surnamestart Yau\surnameend} (\bibinfo{year}{2021}):
  \emph{\bibinfo{title}{2-dimensional Categories}}.
\newblock \bibinfo{publisher}{Oxford University Press, USA},
  \doi{10.1093/oso/9780198871378.001.0001}.

\bibitemdeclare{book}{essentials}
\bibitem{essentials}
\bibinfo{author}{Kevin \surnamestart Leyton-Brown\surnameend} \&
  \bibinfo{author}{Yoav \surnamestart Shoham\surnameend}
  (\bibinfo{year}{2008}): \emph{\bibinfo{title}{Essentials of Game Theory: A
  Concise Multidisciplinary Introduction}}.
\newblock \bibinfo{publisher}{Morgan \& Claypool},
  \doi{10.1007/978-3-031-01545-8}.

\bibitemdeclare{article}{libkind_operadic_2022}
\bibitem{libkind_operadic_2022}
\bibinfo{author}{Sophie \surnamestart Libkind\surnameend},
  \bibinfo{author}{Andrew \surnamestart Baas\surnameend}, \bibinfo{author}{Evan
  \surnamestart Patterson\surnameend} \& \bibinfo{author}{James \surnamestart
  Fairbanks\surnameend} (\bibinfo{year}{2022}): \emph{\bibinfo{title}{Operadic
  {Modeling} of {Dynamical} {Systems}: {Mathematics} and {Computation}}}.
\newblock {\slshape \bibinfo{journal}{Electronic Proceedings in Theoretical
  Computer Science}} \bibinfo{volume}{372}, pp. \bibinfo{pages}{192--206},
  \doi{10.4204/EPTCS.372.14}.
\newblock \urlprefix\url{http://arxiv.org/abs/2105.12282}.
\newblock \bibinfo{note}{ArXiv:2105.12282 [math]}.

\bibitemdeclare{article}{myers_double_2021}
\bibitem{myers_double_2021}
\bibinfo{author}{David~Jaz \surnamestart Myers\surnameend}
  (\bibinfo{year}{2021}): \emph{\bibinfo{title}{Double {Categories} of {Open}
  {Dynamical} {Systems} ({Extended} {Abstract})}}.
\newblock {\slshape \bibinfo{journal}{Electronic Proceedings in Theoretical
  Computer Science}} \bibinfo{volume}{333}, pp. \bibinfo{pages}{154--167},
  \doi{10.4204/EPTCS.333.11}.
\newblock \urlprefix\url{http://arxiv.org/abs/2005.05956}.
\newblock \bibinfo{note}{ArXiv:2005.05956 [math]}.

\bibitemdeclare{book}{myers2021book}
\bibitem{myers2021book}
\bibinfo{author}{David~Jaz \surnamestart Myers\surnameend}
  (\bibinfo{year}{2022}): \emph{\bibinfo{title}{Categorical Systems Theory}}.
\newblock \urlprefix\url{http://davidjaz.com/Papers/DynamicalBook.pdf}.
\newblock \bibinfo{note}{(Work in progress)}.

\bibitemdeclare{inproceedings}{schaefer_competitive_2019}
\bibitem{schaefer_competitive_2019}
\bibinfo{author}{Florian \surnamestart Schaefer\surnameend} \&
  \bibinfo{author}{Anima \surnamestart Anandkumar\surnameend}
  (\bibinfo{year}{2019}): \emph{\bibinfo{title}{Competitive {Gradient}
  {Descent}}}.
\newblock In: {\slshape \bibinfo{booktitle}{Advances in {Neural} {Information}
  {Processing} {Systems}}}, \bibinfo{volume}{32}, \bibinfo{publisher}{Curran
  Associates, Inc.}
\newblock
  \urlprefix\url{https://proceedings.neurips.cc/paper/2019/hash/56c51a39a7c77d8084838cc920585bd0-Abstract.html}.

\bibitemdeclare{incollection}{shulman2021homotopy}
\bibitem{shulman2021homotopy}
\bibinfo{author}{Michael \surnamestart Shulman\surnameend}
  (\bibinfo{year}{2021}): \emph{\bibinfo{title}{Homotopy Type Theory: The Logic
  of Space}}.
\newblock \bibinfo{publisher}{Cambridge University Press}, p.
  \bibinfo{pages}{322}, \doi{10.1017/9781108854429.009}.

\bibitemdeclare{unpublished}{smithe2021compositional}
\bibitem{smithe2021compositional}
\bibinfo{author}{Toby St~Clere \surnamestart Smithe\surnameend}
  (\bibinfo{year}{2021}): \emph{\bibinfo{title}{Compositional Active Inference
  I: Bayesian Lenses. Statistical Games}}.
\newblock \urlprefix\url{https://arxiv.org/abs/2109.04461}.

\bibitemdeclare{unpublished}{spivak2019generalized}
\bibitem{spivak2019generalized}
\bibinfo{author}{David~I \surnamestart Spivak\surnameend}
  (\bibinfo{year}{2019}): \emph{\bibinfo{title}{Generalized Lens Categories via
  functors $F:\mathcal C^{\rm op} \to \mathsf{Cat}$}}.
\newblock \urlprefix\url{https://arxiv.org/abs/1908.02202}.

\bibitemdeclare{article}{spivak1973comprehensive}
\bibitem{spivak1973comprehensive}
\bibinfo{author}{Michael \surnamestart Spivak\surnameend}
  (\bibinfo{year}{1973}): \emph{\bibinfo{title}{A comprehensive introduction to
  differential geometry}}.
\newblock {\slshape \bibinfo{journal}{Bulletins of the American Mathematical
  Society}} \bibinfo{volume}{79}, pp. \bibinfo{pages}{303--306},
  \doi{10.1090/S0002-9904-1973-13149-0}.

\bibitemdeclare{unpublished}{videla2022lenses}
\bibitem{videla2022lenses}
\bibinfo{author}{Andre \surnamestart Videla\surnameend} \&
  \bibinfo{author}{Matteo \surnamestart Capucci\surnameend}
  (\bibinfo{year}{2022}): \emph{\bibinfo{title}{Lenses for Composable
  Servers}}.
\newblock \urlprefix\url{https://arxiv.org/abs/2203.15633}.

\end{thebibliography}

\end{document}